\documentstyle [12pt] {article}

\renewcommand{\theequation}{\arabic{section}.\arabic{equation}}
\makeatletter\@addtoreset{equation}{section}\makeatother

\begin{document}
\begin{titlepage}
\rightline{LA-UR-92-3856}
\rightline{November, 1992} \vskip .1in

\begin{center}
{\LARGE Particle production in the \\ 
central rapidity region}\\
\ \\\ \\
{\large
F. Cooper,$^1$  J. M. Eisenberg,$^2$
Y. Kluger,$^{1,2}$\\
E. Mottola,$^1$ and B. Svetitsky$^2$\\
\ \\}
$^1$Theoretical Division and\\
Center for Nonlinear Studies\\
Los Alamos National Laboratory\\
Los Alamos, New Mexico 87545 USA\\
\ \\
$^2$School of Physics and Astronomy\\
Raymond and Beverly Sackler Faculty of Exact Sciences\\
Tel Aviv University, 69978 Tel Aviv, Israel
\ \\ \ \\

{\em ABSTRACT}
\end{center}
\quotation\noindent

We study pair production from a strong electric field in boost-invariant
coordinates as a simple model for the central rapidity region of a heavy-ion
collision. We derive and solve the renormalized equations for the time
evolution of the mean electric field and current of the produced particles,
when the field is taken to be a function only of the fluid proper time $\tau =
\sqrt{t^2-z^2}$. We find that a relativistic transport theory with a Schwinger
source term modified to take Pauli blocking (or Bose enhancement) into account
gives a good description of the numerical solution to the field equations. We
also compute the renormalized energy-momentum tensor of the produced particles
and compare the effective pressure, energy and entropy density to that expected
from hydrodynamic models of energy and momentum flow of the plasma.
\endquotation
\end{titlepage}


\section{Introduction}

A popular theoretical picture of high-energy heavy-ion collisions begins
with the creation of a flux tube containing a strong color electric field
\cite{Biro}. The field energy is converted into particles as $q \bar{q}$
pairs and gluons are created by the Schwinger tunneling
mechanism \cite{Sauter,HE,Schwinger51}. The transition from this quantum
tunneling stage to a later hydrodynamic stage has previously been described
phenomenologically using a kinetic theory model in which a relativistic
Boltzmann equation is coupled to a simple Schwinger source term
\cite{BC86,BC88,Kajantie,Gatoff}. Such a model requires justification, as does
the use of Schwinger's formula in the case of an electric field which is
changing rapidly because of screening by the produced particles. Our aim in
this paper is to present a completely field-theoretic treatment of the
electrodynamic initial-value problem which exhibits the decay of the electric
field and subsequent plasma oscillations. This approach allows direct
calculation of the spectrum of produced particles from first principles and
comparison of the results with more phenomenological hydrodynamic models of the
plasma.

Although our approach is relevant to heavy-ion collisions at best only during
the period when the produced partons can be treated as almost free, the details
of hadronization are not expected to affect the average flow of energy and
momentum. Hence information obtained about energy flow in the weak coupling
phase where our methods apply is translated to energy and momentum eventually
deposited in the detector. Using further hadronization assumptions one can
relate our results to the particle spectrum of the outgoing particles.

Recently \cite{CM,PRL,PRD,Revs} we have presented a practical renormalization
scheme appropriate for initial-value and quantum back reaction problems
involving the production of charged pairs of bosons or fermions by a strong
electric field. In those papers the electric field is restricted to be
spatially homogeneous, so that all physical quantities are functions of time
alone. The method used to identify the divergences is to perform an adiabatic
expansion of the equations of motion for the time-dependent mode functions. The
divergence in the expectation value of the current comes from the first few
terms in the adiabatic expansion and can be isolated and identified as the
usual coupling-constant counter term.  In this manner we were able to construct
finite equations for the process of pair production from a spatially
homogeneous electric field, and to consider the back reaction that this pair
production has on the time evolution of the electric field.

In heavy-ion collisions one is clearly dealing with a situation that is not
spatially homogeneous. However, particle production in the central rapidity
region can be modeled as an inside-outside cascade which is symmetric under
longitudinal boosts and thus produces a plateau in the particle rapidity
distributions \cite{Feynman,BjCC,CKS,Bjorken,Lund}. This boost invariance also
emerges dynamically in Landau's hydrodynamical model \cite{Landau} and forms an
essential kinematic ingredient in the subsequent models of Cooper, Frye, and
Schonberg \cite{CFS} and of Bjorken \cite{Bjorken}. The flux-tube model of Low
\cite{Low} and Nussinov \cite{Nussinov} incorporates this invariance naturally.

Hence, kinematical considerations constrain the spatial inhomogeneity in the
central rapidity region to a form that again allows an adiabatic expansion in a
single variable, the fluid proper time $\tau = \sqrt{t^2-z^2}$\@. In Landau's
hydrodynamical model one finds that after a short time $\tau_0$ the
energy-momentum tensor in a comoving frame is a function only of $\tau$\@.
This is the essential assumption we make that allows us to apply the methods of
\cite{CM,PRL,PRD,Revs} to the heavy-ion collision problem.

In our approach the initial conditions of the fields are specified at
$\tau=\tau_0$, that is, on a hyperbola of constant proper time. The comoving
energy density is a function of $\tau$ only.\footnote{Following
the usage in hydrodynamics we shall continue to refer to the
coordinates $\tau$ and $\eta = {\frac {1}{2}} \ln\left(\frac{t+z}{t-z}
\right)$ as the fluid proper time and the fluid rapidity, respectively, or
simply as comoving coordinates.} Then the electric field must also
be a function solely of $\tau$\@. We can apply the adiabatic regularization
method to identify and remove the divergences.
The simplification introduced by the boost symmetry allows us to
study the renormalization of this inhomogeneous initial value
problem.

A remarkable feature of Landau's model is the appearance of a scaling solution
in which $v_z =z/t$\@. Alternatively, requiring invariance under longitudinal
boosts \cite{Bjorken,CFS} also leads naturally to a scaling solution. In
models based on Landau's ideas, one also assumes that it is possible to
determine the particle spectra from the hydrodynamical flow of energy-momentum
by identifying particle velocities with hydrodynamical velocities.
Using our methods, we can assess the validity of these assumptions.
We calculate numerically the evolution of the electric field in
$\tau$ and study the accompanying production of pairs. As it turns out, the
Boltzmann equation, when modified to reflect quantum statistics correctly, does
very well at reproducing the gross features of the field-theoretic solution.
This is not too surprising, since our field equations are mean field equations
(formally derived in the large-$N$ limit), and hence are semiclassical in
nature. The field theoretic treatment presented here is the first term in a
systematic $1/N$ expansion in the number of parton species or quark flavors.
The next order in the expansion contains dynamical gauge fields as well as
charged particles in the background classical field. Thus in the next order one
can study equilibration due to scattering; one could also calculate, for
example, lepton production and correlations in the evolving plasma from first
principles. These systematic corrections require the use of the
Schwinger-Keldysh formalism \cite{SchKel} and will be discussed elsewhere.

The outline of the paper is as follows. In Section 2 we formulate
electrodynamics in the semiclassical limit in the $(\tau, \eta )$ coordinate
system corresponding to the hydrodynamical scaling variables. This curvilinear
coordinate system requires some formalism borrowed from the literature of
quantum fields in curved spaces, which we review for the benefit of the reader
unfamiliar with the subject. In Section 3 we perform the renormalization of the
current using the adiabatic method of our previous papers. This is needed as
the source term for a finite Maxwell equation of the particle back reaction on
the electric field. Section 4 is devoted to the renormalization of the
energy-momentum tensor of the produced pairs in the comoving frame. We also
discuss there the relationship to the effective hydrodynamic point of view. The
detailed results of numerical calculations in the (1+1)-dimensional case for
both charged scalars and fermions are presented in Section 5.

The paper contains three appendices. In Appendix A we present the necessary
formulae used in computing the particle spectrum from the time evolution of the
field modes. In Appendix B we prove that, for the boost-invariant kinematics of
this problem, the distribution of particles in fluid rapidity is the same as
the distribution of particles in particle rapidity. In Appendix C we
reformulate the problem in the conformal time coordinate which turns out to be
somewhat more convenient for actual numerical methods.

\section {Electrodynamics in comoving coordinates}

\subsection{Scalars}
We consider first the electrodynamics of spin-0 bosons. We shall use the metric
convention $(-+++)$ which is commonly used in the curved-space literature.
The action in general curvilinear coordinates with metric $g_{\mu \nu}$ is
\begin{equation}
S = \int d^{4}x \,\sqrt{-g} \,\left[ - g^{\mu \nu}
(\nabla_{\mu}  \phi)^{\dagger} \nabla_{\nu} \phi -   m^{2}
\phi^{\dagger} \phi
-\frac{ 1}{4}  g^{\mu \rho}
g^{\nu \sigma} F_{\mu \nu} F_{\rho \sigma} \right]\,,
 \label{boost_S}
\end{equation}
where
\begin{equation}
       \nabla_{\mu}\phi = (\partial_{\mu} - i e A_{\mu})\phi\,,\quad
F_{\mu \nu} = \partial_{\mu} A_{\nu} -  \partial_{\nu} A_{\mu}\,.
\label{boost_F}
\end{equation}
We use Greek indices for curvilinear coordinates and Latin indices for flat
Minkowski coordinates.

To express the boost invariance of the system it is useful to introduce the
light-cone variables $\tau$ and $\eta$, which will be identified later with
fluid proper time and rapidity . These coordinates are defined in terms of the
ordinary lab-frame Minkowski time $t$ and coordinate along the beam direction
$z$ by
\begin{equation}
       z= \tau \sinh \eta  \quad ,\quad t= \tau \cosh \eta \,.
\label{boost_tz.tau.eta}
\end{equation}
 The Minkowski line element in these coordinates has the form
\begin{equation}
{ds^2} = {- d \tau^2 + dx^2   +dy^2 +{\tau}^2 {d \eta}^2 }\,.
\label{boost_line_element}
\end{equation}
Hence the metric tensor is given by
\begin{equation}
 g_{\mu \nu} = {\rm diag} (-1, 1, 1, \tau^2).
\label{boost_metric}
\end{equation}
with its inverse determined from $g^{\mu \nu} g_{\nu \rho} =
\delta^{\mu}_{\rho}$
This metric is a special case of the Kasner metric \cite{Birrell_Davies}.

For our future use we introduce the vierbein $V^a_{\mu}$ which transforms the
curvilinear coordinates to Minkowski coordinates,
\begin{equation}
g_{\mu \nu} = V^{a}_{\mu}  V^{b}_{\nu} \eta_{ab}\,,
\label{boost_gmunuD}
\end{equation}
where
$\eta_{ab}= {\rm diag} \{-1,1,1,1\}$\@ is the flat Minkowski metric.
A convenient choice of the vierbein for the metric (\ref{boost_metric}) for our
problem is
\begin{equation}
   V_{\mu}^{a}  = {\rm diag}\{1,1,1,\tau \}
\end{equation}
so that
\begin{equation}
V^\mu_a= {\rm diag}\left\{1,1,1,{\frac {1}{\tau}}\right\}\,.
\label{boost_vierbein}
\end{equation}
Thus the determinant of the metric tensor is given by
\begin{equation}
\det V = \sqrt{-g} = \tau\,.
\label{boost_detV}
\end{equation}

The Klein-Gordon  equation is
\begin{equation}
\frac {1}{\sqrt{-g}}\nabla_{\mu} (g^{\mu \nu}\sqrt{-g})
\nabla_{\nu} \phi - m
^2 \phi  = 0\,,
\label{boost_KG}\end{equation}
and Maxwell's equations read
\begin{equation}
\frac {1}{\sqrt{-g}}\partial_{\nu} (\sqrt{-g} F^{\mu \nu}) = j^{\mu}\,,
\label{boost_Max}
\end{equation}
where
\begin{equation}
j^{\mu} = {\cal C} \{ -ie[\phi^{\dagger} \partial^{\mu} \phi -(\partial^{\mu}
\phi^{\dagger})\phi] - 2 e^2 A^{\mu} (\phi^{\dagger} \phi) \}\, .
\label{boost_current}
\end{equation}
Here $\cal C$ denotes the operation of charge symmetrization as
discussed in \cite{CM}.

We are interested in the case where the electric field is in the
$z$  direction and is a function of $\tau$ only.
In Minkowski coordinates the only nonvanishing components of the
electromagnetic field tensor $ F_{ab}$ are
\begin{equation}
  F_{zt} = - F_{tz} \equiv E(\tau ) \,.
\label{boost_Fzt}
\end{equation}
In the curvilinear coordinate system we have
\begin{equation}
F_{\eta \tau} = - \frac {dA_{\eta}(\tau)}{d\tau}\,,
 \label{boost_Feta}
\end{equation}
where we have chosen the gauge  $A_{\tau} = 0$ so that
the only nonvanishing component of $A_{\mu}$ is
$A_{\eta} (\tau) \equiv A$\@.
Using the relationship between the two coordinate systems we find that
\begin{equation}
E(\tau) = \frac{F_{\eta \tau}}{\tau} = - {1\over \tau} {dA \over d\tau}\,.
\label{boost_Etau}
\end{equation}

In this gauge and coordinates the Klein-Gordon equation becomes
\begin{equation}
\left(\partial^2 _{\tau} +\frac{1}{\tau}\partial_{\tau} -\frac{1}{\tau^2}
{(\partial_{\eta}
-ieA(\tau))}^2  - {\partial_x}^2-{\partial_y}^2  +
 m^2  \right) \phi(\tau,\eta,x,y) = 0\,.
\label{boost_KGs}
\end{equation}
In order to remove first derivatives with respect to $\tau$, we define a
rescaled field $\chi$  by
\begin{equation}
\phi =\frac {1}{\sqrt{\tau}} \chi \,.
\label{boost_chi}
\end{equation}
The Klein-Gordon equation for $ \chi$ is then
\begin{equation}
\left(\partial^2 _{\tau} - \frac{1}{\tau^2}\left[ {(\partial_{\eta}
-ieA(\tau))}^2 - {\frac {1}{4}}\right] - {\partial_x}^2-{\partial_y}^2  +
 m^2  \right) \chi(\tau,\eta,x,y) = 0\,.
\label{boost_KGchi}
\end{equation}
We are interested in the solution of this field equation, where
$A$ is regarded as a classical field determined by
the  expectation value of the Maxwell equation (\ref{boost_Max}).
This approximation ignores processes with photon propagators, and
can be shown \cite{CM} to be
the leading order in  a  large-$N$ expansion, where $N$
is the number of flavors of the charged scalar field.

These equations are to be solved subject to initial conditions at
$\tau = \tau_{0}$\@.  We need to specify the
initial value of the electric field and the density matrix describing
the initial state of the charged scalar field. For the problem at
hand it is sufficient to describe the charged scalar field by the
particle-number density and pair-correlation density with respect to an
adiabatic vacuum state [see (\ref{boost_NF}) below].

In the gauge we have chosen there is homogeneity in $\eta$ as well as in
the directions $x$ and $y$\@.
This allows us to introduce a Fourier decomposition for the
quantum field operator  $\chi$ at proper time $\tau$,
\begin{equation}
\chi(\tau,\eta,{\bf x}_{\perp}) = \int [d{\bf k}]
\left[f_{\bf k}(\tau) a_{\bf k} e^{i {{\bf k} \cdot{\bf x}}} +
f^* _{\bf{-k}}(\tau) b^{\dagger}_{\bf k} e^{-i  {{\bf k} \cdot
{\bf x}}}\right]\,, \label{boost_field}
\end{equation}
where
\begin{eqnarray}
 [d{\bf k}] &=& \frac{dk_{\eta} d^{2}{\bf k}_{\perp}}{
(2\pi)^{3}}\,,\nonumber \\ {{\bf k} \cdot {\bf x}} &=& k_{\eta}\eta +
{\bf k}_{\perp}  \cdot {\bf x}_{\perp}\,,\nonumber \\
{\bf k}_{\perp}&\equiv & (k_x,k_y)\,, \quad {\bf x}_{\perp}\equiv (x,y)\,.
\label{boost_momentum}
\end{eqnarray}
The modes $f_{\bf k}$ satisfy the equation
\begin{equation}
{d^2 f_{\bf k} \over d \tau^2}
+ \omega_{\bf k}^2(\tau) f_{\bf k} = 0\,,
\label{boost_mode_eq_b}
\end{equation}
with
\begin{eqnarray}
\omega_{\bf k}^2 (\tau)&\equiv&
\pi_{\eta}^2(\tau) +{\bf k}_{\perp}^2+m^{2}+\frac{1}{4 \tau^2}\,,\nonumber \\
\pi_{\eta}(\tau) &\equiv& \frac{k_{\eta}-eA}{\tau}\,.
\label{boost_omega_b}
\end{eqnarray}
We quantize the matter field  by imposing commutation relations at equal
$\tau$,
\begin{eqnarray}
\left[\phi(\tau,\eta,{\bf x}_{\perp}), \frac{\partial \phi^{\dagger}}
{\partial \tau} (\tau,\eta',{\bf x}'_{\perp}) \right ]
= \frac{i\delta (\eta-\eta')
\delta^2 ({\bf x}_{\perp}-{\bf x}'_{\perp})}{\tau}\,.
\label{boost_commutationa}
\end{eqnarray}
Demanding that the usual commutation relations obtain for the creation and
annihilation operators,
\begin{equation}
[b_{\bf{ k}}, b^{\dagger}_{\bf k}] =[a_{\bf k}, a^{\dagger}_
{\bf k}] = (2 \pi)^d \delta^d({\bf k}-{\bf k}')\,,
\label{boost_commutationb}
\end{equation}
in $d$ spatial dimensions, we find that $f_{\bf k}$ must satisfy the condition
\begin{equation}
f_{\bf k}(\tau) \partial_{\tau} f_{\bf k}^{\ast} (\tau) -
f_{\bf k}^{\ast}(\tau) \partial_{\tau}
f_{\bf k}(\tau) = i \,.
\label{boost_wronskian}
\end{equation}
This latter condition is satisfied by a WKB-like parametrization
of $f_{\bf k}$,
\begin{equation}
f_{\bf k} = {{e^{-i \int_0^\tau \Omega_{\bf k}(\tau')
d\tau'}}
\over {(2 \Omega_{\bf k}(\tau))^{1/2}}}
 \equiv {{e^{-iy_{\bf k}(\tau)}} \over {(2 \Omega_{\bf
k}(\tau))^{1/2}}}\,.
\label{boost_fansatzb}
\end{equation}
Because of (\ref{boost_mode_eq_b}), $\Omega_{\bf k}$ must satisfy
the same differential equation as appears in our
previous  papers \cite{CM,PRL,PRD,Revs},
\begin{equation}
{1 \over 2} \frac {\ddot\Omega_{\bf k}}{\Omega_{\bf k}} - {3 \over 4}
{\dot\Omega_{\bf k}^2 \over \Omega_{\bf k}^2}
+ \Omega_{\bf k}^2 = \omega_{\bf k}^2 .
\label{boost_WKBb1}
\end{equation}
The dot denotes differentiation with respect to $\tau$\@.

The only nontrivial Maxwell equation in $A_{\tau} = 0$ gauge is
\begin{equation}
{1 \over \tau} {d \over d\tau} \left[{1 \over \tau}
{d \over d\tau}A(\tau)\right ]= \langle j^{\eta} \rangle\, .
\label{boost_Maxb}
\end{equation}
In terms of the charge densities $N_+$ and~$N_-$
and the correlation pair density $F$, we can write the Maxwell
equation as
\begin{equation}
 - \tau {\frac {d E} {d \tau}} = e \int [d{\bf k}]
\frac{\pi_{\eta}}
 {\Omega_{\bf k} } [1+2N({\bf k})+2F({\bf k})\cos(2y_{\bf k})]\,.
\label{boost_maxmode1}
\end{equation}
The structure of (\ref{boost_maxmode1}) is similar to that of
the equation found for the homogeneous problem \cite{CM,PRL}.
We have used the definitions
\begin{eqnarray}
\langle a^\dagger_{{\bf k}^\prime} a_{\bf k} \rangle &=&
(2\pi)^d \delta^d({\bf k}-{\bf k}^\prime)
N_{+}({\bf k})\,,\nonumber\\
\langle b^\dagger_{-{\bf k}^\prime} b_{-{\bf k}} \rangle &=&
(2\pi)^d \delta^d({\bf k}-{\bf k}^\prime)
N_{-}({\bf k})\,,\nonumber \\
\langle b_{-{\bf k}^\prime} a_{{\bf k}} \rangle &=&
(2\pi)^d \delta^d({\bf k}-{\bf k}^\prime)
F({\bf k})\,.
\label{boost_NF}
\end{eqnarray}
Note that we have taken $N_{+}({\bf k})=N_{-}({\bf k})=N({\bf k})$
since the current component $j^{\tau}$ vanishes due to the Maxwell equation
(Gauss's law),
\begin{equation}
j^{\tau}=e\int \frac{[d{\bf k}]}{\tau}\left [N_{+}({\bf k})-N_{-}({\bf k})
\right]=\frac {1}{\sqrt{-g}}\partial_{\eta} (\sqrt{-g} F^{\eta \tau})  =0\,.
\end{equation}

\subsection{Fermions}

Let us now turn to the same problem in Dirac electrodynamics.
The lagrangian density for QED in curvilinear
coordinates (found, for example, in \cite{Birrell_Davies})
gives rise to the action
\begin{equation}
S= \int d^{d + 1}x \, ({\rm{det}}\, V) \left[ {\frac{-i}{2}}
\bar {\Psi} \tilde{\gamma}^{\mu}
\nabla_{\mu} \Psi+ {\frac{i}{2}} (\nabla^{\dag}_{\mu}\bar {\Psi} )
\tilde{\gamma}^{\mu} \Psi  -i m \bar {\Psi}\Psi-
\frac {1}{4}F_{\mu \nu}F^{\mu \nu} \right],
\label{boost_Sf}
\end{equation}
where \cite{Weinberg}
\begin {equation}
\nabla_{\mu} \Psi \equiv (\partial_{\mu} + \Gamma_{\mu} -ieA_{\mu}) \Psi
\end{equation}
and the spin connection $\Gamma_{\mu}$ is given by
\begin {equation}
\Gamma_{\mu}=\frac{1}{2}\Sigma^{ab}V_{a {\nu}}(\partial_{\mu} V_b^{\nu} +
\Gamma^{\nu}_{\mu \lambda} V^{\lambda}_b ) \,,\qquad
\Sigma^{ab}=\frac{1}{4}[\gamma^a,\gamma^b]\,,
\label{boost_nabla}
\end{equation}
with $\Gamma^{\nu}_{\mu \lambda}$ the usual Christoffel symbol.
We find that in our case (see \cite{Parker})
\begin{eqnarray}
\Gamma_{\tau}&=&\Gamma_x=\Gamma_y=0 \nonumber \\
\Gamma_{\eta}&=&-\frac{1}{2}\gamma^0 \gamma^3 .
\label{boost_Gamma}
\end{eqnarray}
The coordinate dependent gamma matrices
$\tilde{\gamma}^{\mu}$ are obtained from the usual Dirac gamma matrices
$\gamma^{a}$
via
\begin{equation}
 \tilde{\gamma}^{\mu} = \gamma^{a} V_{a}^{\mu}(x)\ .
\end{equation}
The coordinate independent Dirac matrices satisfy
\begin{equation}
\{\gamma^{\alpha},\gamma^{\beta}\} = 2 \eta^{\alpha \beta}.
\label{boost_gammaD}
\end{equation}

{}From the action (\ref{boost_Sf}) we obtain the Heisenberg field
equation for the fermions,
\begin{equation}
\left( \tilde{\gamma}^{\mu}\nabla_{\mu} + m \right) \Psi=0\,,
\end{equation}
which takes the form
\begin{equation}
\left[ \gamma^0 \left(\partial_\tau+\frac{1}{2\tau}\right)
+{\bf \gamma}_\perp\cdot \partial_\perp
+ \frac{\gamma^3}{\tau}(\partial_\eta -ieA_\eta)+ m \right] \Psi =0\,,
\label{boost_Dirac}
\end{equation}
Variation of $S$ with respect to $A_{\mu}$ yields the semiclassical Maxwell
equation
\begin{eqnarray}
{\frac {1} {\sqrt{-g}}} \, \partial_{\nu} \left(\sqrt{-g}F^{\mu \nu}
\right) =
\langle j^{\mu} \rangle
= - {\frac{e}{2}}\left \langle \left[\bar \Psi, \tilde {\gamma} ^{\mu} \Psi
\right]\right \rangle  .
\label{boost_MaxD1}
\end{eqnarray}
If the electric field is in the $z$~direction and a function of $\tau$ only,
we find that the only nontrivial Maxwell equation is
\begin{equation}
\frac{1}{\tau} {\frac {dE(\tau)} {d\tau}} =  {\frac{e}{2}} \left \langle \left[
\bar{\Psi}, \tilde {\gamma}^{\eta} \Psi \right] \right \rangle
=\frac{e}{2\tau} \left \langle \left[ \Psi^{\dagger}, \gamma^0\gamma^3 \Psi
\right] \right \rangle .
\label{boost_MaxD2}
\end{equation}

We expand the fermion field in terms of Fourier modes at fixed proper time
$\tau$,
\begin{equation}
\Psi (x) = \int [d{\bf k}] \sum_{s}[b_{s}({\bf k})
\psi^{+}_{{\bf k}s}(\tau)
 e^{i k \eta} e^{ i {\bf{p}} \cdot {\bf x}}
+d_{s}^{\dagger}({\bf{-k}}) \psi^{-}_{{\bf{-k}}s}(\tau)
e^{-i k \eta} e^{ - i {\bf{p}} \cdot {\bf x}}  ].
\label{boost_fieldD}
\end{equation}
The $\psi^{\pm}_{{\bf k}s}$ then obey
\begin{eqnarray}
 \left[\gamma^{0} \left({d\over d \tau}+\frac{1}{2\tau}\right)
+ i \gamma_{\bf{{\perp}}}\cdot{\bf{k_{\perp}}}
+i {\gamma^{3}} \pi_{\eta}
 + m\right]\psi^{\pm}_{{\bf k}s}(\tau) = 0,
\label{boost_mode_eq_D}\end{eqnarray}
where $\pi_{\eta}$ has been defined previously in (\ref{boost_omega_b}).
The superscript $\pm$ refers to positive- or negative-energy solutions
with respect to the adiabatic vacuum at $\tau= \tau_{0}$\@.
Following \cite{PRD}, we square the Dirac equation by introducing
\begin{equation}
\psi^{\pm}_{{\bf k}s} = \left[-\gamma^{0}\left( {d \over d\tau}
+\frac{1}{2\tau}\right)
- i \gamma_{\bf{{\perp}}}\cdot {\bf{k_{\perp}}}
-i \gamma^{3} \pi_{\eta}+ m\right] \chi_{s} {f^{\pm}_{{\bf k}s} \over {\sqrt
\tau}}
\, . \label{boost_psi_g}
\end{equation}
The spinors $\chi_s$ are chosen to be eigenspinors of $\gamma^0\gamma^3$,
\begin{equation}
\gamma^{0}\gamma^{3}\chi_{s} = \lambda_{s} \chi_{s}\,
\end{equation}
with $\lambda_{s}=1$ for $s=1,2$ and $\lambda_{s}=-1$ for $s=3,4$.
They are normalized,
\begin{equation}
\chi^{\dag}_{r} \chi_{s} = 2\delta_{rs} \,.
\label{boost_e.v}
\end{equation}
The sets $s={1,2}$ and $s={3,4}$ are two different complete sets of
linearly independent solutions of the Dirac equation (see \cite{PRD}).
Inserting (\ref{boost_psi_g}) into the Dirac equation
(\ref{boost_mode_eq_D})  we obtain the quadratic mode equation
\begin{equation}
\left( \frac{d^2}{d \tau^2}+
\omega_{\bf k}^2 -i \lambda_{s} \dot{\pi}_{\eta} \right )
f^{\pm}_{{\bf k}s}(\tau) = 0,
\label{boost_modef_D}
\end{equation}
where now
\begin{equation}
 \omega_{\bf k}^2= \pi_{\eta}^2 +{\bf k}_{\perp} ^2 +m^2 .
\label{boost_omega_D}
\end{equation}

If the canonical anti-commutation relations are imposed on the Fock space mode
operators, then the $\psi^{\pm}_{{\bf k}s}$ must obey the orthonormality
relations
\begin{eqnarray}
\psi^{-\dag}_{{\bf k}r} \psi^{+}_{{\bf k}s} &=&
\psi^{+\dag}_{{\bf k}r} \psi^{-}_{{\bf k}s}=0\,, \nonumber \\
\psi^{+\dag}_{{\bf k}r} \psi^{+}_{{\bf k}s} &=&
\psi^{-\dag}_{{\bf k}r}
 \psi^{-}_{{\bf k}s}=\frac{\delta_{r,s}}{\tau}\,,
\label{boost_ortho_D}
\end{eqnarray}
where $r, s=1,2$ or $3,4$\@.
Using the orthonormality relations and (\ref{boost_psi_g}) we find,
for a given $\bf k$ and $s$,
\begin{equation}
\omega^2  {f^{\ast \alpha }}
f^{\beta}+
{{\dot{f}}^{\ast \alpha }}
\dot{f}^{\beta}-i\lambda_{s}
\pi \left ({f^{\ast \alpha }}
\dot{f}^{\beta}-
{\dot{f}^{\ast \alpha }} f^{\beta}\right )
= {\frac{1}{2}} \delta^{\alpha \beta}
\label{boost_norm_D}
\end{equation}
where $ \alpha , \beta = \pm$ refer to the positive and negative energy
solutions.
Notice that from the off-diagonal relationship ($\alpha \ne \beta$)
we can express the negative-energy solutions in terms of
the positive-energy ones. This fact we shall use repeatedly in what follows.

Now we parametrize the positive-energy solutions $f^{+}_{{\bf k}s}$
in the same manner as in Eq.~(3.1) of Ref.~\cite{PRD},
\begin{eqnarray}
f_{{\bf k}s}^+ (\tau) = N_{{\bf k}s}
 \frac {1}{\sqrt{2\Omega_{{\bf k}s}}} \exp\left \{ \int_{0}^{\tau}
\left ( -i\Omega_{{\bf k}s} (\tau')
-\lambda_{s}  \frac {{\dot{\pi}
_{\eta}}(\tau')}
{2\Omega_{{\bf k}s} (\tau')}
\right )
d\tau'\right \}  ,
\label{boost_ansatz_D}
\end{eqnarray}
where $\Omega_{{\bf k}s}$ obeys the real equation
\begin{equation}
{1 \over 2} \frac {\ddot\Omega_{{\bf k}s}}{\Omega_{{\bf k}s}} - {3 \over 4}
{\dot\Omega_{{\bf k}s}^2 \over \Omega_{{\bf k}s}^2}
+{\lambda_{s} \over 2} \frac {\ddot\pi_{\eta}}{\Omega_{{\bf k}s}}
 -{1 \over 4}
{\dot\pi_{\eta}^2 \over \Omega_{{\bf k}s}^2} - \lambda_{s}
{\frac{\dot{\pi}_{\eta}\dot{\Omega}_{{\bf k}s}} {\Omega_{{\bf k}s}^2}}
=\omega_{\bf k}^2(\tau) - \Omega_{{\bf k}s}^2\,.
\label{boost_WKB_D}
\end{equation}

Returning to the Maxwell equation and following steps (2.27)--(2.30) of
Ref.~\cite{PRD} we obtain
\begin{equation}
\frac{1}{\tau}\frac{dE(\tau)}{d\tau} = -\frac{2e}{\tau^2} \sum_{s=1}^{4}\int
 [d{\bf k}]
({\bf k}^2_{\perp} +m^2)
\lambda_{s}\vert f_{{\bf k}s}^{+}\vert ^2 ,
\label{boost_MaxD3}
\end{equation}
where we have taken the particle number $N ({\bf k} s)$ and pair correlation
density $F ({\bf k} s)$ defined by the analogs of eqs. (\ref{boost_NF}) equal
to zero for simplicity.

Using the normalization conditions (\ref{boost_ortho_D})--(\ref{boost_norm_D})
we may express the mode functions and current in terms of the generalized
frequency functions $\Omega_{{\bf k}s}$,
\begin{equation}
2 \vert f_{{\bf k}s}^{+}\vert ^2 = \left[ \omega_{\bf k}^2 +
\Omega_{{\bf k}s}^2 +
\left ( \frac{ \dot{\Omega}_{{\bf k}s} +\lambda_s \dot{\pi}_{\eta} }
{2\Omega_{{\bf k}s}} \right )^2
 + 2\lambda_s\pi_{\eta}\Omega_{{\bf k}s} \right ]^{-1}\,.
\label{boost_MaxD4}
\end{equation}
[See Eq. (3.7) of Ref.~\cite{PRD}.]

\section{Renormalization}
\subsection{Scalars}

The renormalization of the equations of the last section
is straightforward, and is accomplished by analyzing
the divergences in  an adiabatic
expansion of the differential equation for $\Omega_{\bf k}(\tau)$\@.
We first present the regularization for the scalar case, where
$\Omega_{\bf k}$ satisfies the differential equation (\ref{boost_WKBb1}).
The divergences of physical
quantities such as the current and the energy-momentum tensor
can be isolated by expanding
$\Omega$  in  an adiabatic expansion.
Up to second order this is given by
\begin{equation}
{1 \over \Omega} = {1 \over \omega} + \left(  {{\ddot \omega}
\over {4 \omega^4}} - {{3 \dot \omega^2} \over {8 \omega^5}} \right)+\cdots
\quad.
\label{boost_Omegainv_ad}
\end{equation}

The unrenormalized Maxwell equation is
\begin{equation}
 - \tau{\frac {d {E}} {d \tau}} = e \int [d{\bf k}]\,
\frac{\pi_{\eta}}{\Omega_{\bf k} (\tau)} \,[1+2N({\bf k})+2F({\bf k})
\cos(2y_{\bf k}(\tau))]\,.
\label{boost_Max_mode2}
\end{equation}
To study its renormalization in $d=3$ spatial dimensions we need to consider
only the vacuum term,
\begin{equation}
-\tau\frac{d E}{d \tau} =e \int [d{\bf k}]\,
\frac{\pi_{\eta}} {\Omega_{\bf k} (\tau)}\,.
\label{boost_Max_vac}
\end{equation}
The adiabatic expansion (\ref{boost_Omegainv_ad}) leads to
\begin{equation}
-\tau\frac{d E}{d \tau} =
e \int [d{\bf k}] \,\frac{(k_{\eta} - eA_{\eta})}{\tau} \,\left[
{1 \over \omega_{\bf k}} + \left({{\ddot \omega_{\bf k}} \over
{4\omega_{\bf k}^4}} - {{3 \dot \omega_{\bf k}^2} \over {8 \omega_{\bf k}^5}}
\right)\right]+ \cdots \quad .
\label{boost_Max_exp1}
\end{equation}
The first term in (\ref{boost_Max_exp1}) is zero by reflection symmetry if we
 choose
fixed integration boundaries for the kinetic momentum
$k_{\eta}-eA_{\eta}$\@.  The only
divergent terms  occur at second order.
(Higher terms in the expansion have more powers of ${\bf k}$ in the
denominator.)
Using (\ref{boost_omega_b}) and reflection symmetry, the right-hand side of
(\ref{boost_Max_exp1}) can be written as
\begin{equation}
e \int [d{\bf k}] \,(k_{\eta}- eA)^{2}
 \left \{{  \frac {e \dot{A}} {\tau^4 \omega_{\bf k}^5}
-\frac{e \ddot{A}}{4
\tau^3 \omega_{\bf k}^5} -\frac{5 e \dot{A}
\left[(k_{\eta}-eA)^2+\frac14\right]}{4 \tau^6 \omega_{\bf k}^7} }
\right\}.
\label{boost_Max_exp2}
\end{equation}

Performing the $k_{\eta}$ and azimuthal angular integrations first we obtain
\begin{equation}
-{e^2 \over 48 \pi^2} \int_0^{\Lambda^2} dk_{\perp}^2 \left\{ {{\ddot A} -
{{\dot A} \over \tau}\over (k_{\perp}^2 + m^2 + {1\over 4 \tau^2})} +{{\dot A}
\over 2 \tau^3 (k_{\perp}^2 + m^2 + {1\over 4 \tau^2})^2} \right\},
\label{boost_inf1}
\end{equation}
where we have inserted a cutoff in the remaining transverse momentum
integration. The logarithmically divergent first term in (\ref{boost_inf1}) is
\begin{eqnarray}
\frac{1}{24 \pi^2} \left (-{\ddot A}+\frac{\dot A}{\tau}\right
)\left [ \ln \left ({{\Lambda}\over m}\right ) - \ln \left (1 + {1\over 4 m^2
\tau^2} \right ) \right ].
\end{eqnarray}
We recognize the cutoff dependent infinite part as
\begin{equation}
e^2  \delta e^2 \tau \frac {d {E}}{d \tau}\,,
\label{boost_inf2}
\end{equation}
where
\begin{equation}
\delta e^2 = (1/24 \pi^2)\ln (\Lambda/ m)
\label{chgeren}
\end{equation}
is the usual one-loop charge renormalization factor in scalar QED.
Defining the renormalized charge via
\begin{equation}
{e_{R}}^{2} = e^{2} (1+e^2 \delta e^2) ^{-1} = e^{2} (1-{e_{R}}^2 \delta
e^2) \,,
\label{boost_eR}
\end{equation}
and using the Ward identity $eE = e_R E_R$, we may absorb the divergence in the
current into the left side of the Maxwell equation to obtain a finite
renormalized equation suitable for numerical integration.

In the $d = 1$  case there is no transverse momentum integration, and the
charge renormalization is finite. The finite coefficient of the combination
$-{\ddot A}+{\dot A}/{\tau}$
is $\tau$-dependent, and is given by
\begin{equation}
\left[12\pi\left(m^2+\frac1{4\tau^2}\right)\right]^{-1}\ .
\label{boost_1Ddeltae}
\end{equation}
The standard result in (1+1) dimensions is
$\delta e^2 =(12\pi m^2)^{-1}$, and this is what is obtained for the
spatially homogeneous problem by our method as well \cite{PRL}.
Thus, absorbing $\delta e^2$ in the renormalization of the charge [see
(\ref{boost_eR})] leaves us with a (finite) $\tau$-dependent coefficient that
is multiplied by the above combination, which appears now on both sides of the
finite Maxwell equation, just as in the three dimensional case. This feature is
peculiar to scalars in the $\tau$ coordinate. The actual numerical solution of
the scalar equations was performed in the conformal time coordinate $u = \ln
(m\tau)$ discussed in Appendix C.

\subsection{Fermions}

We turn to the renormalization problem in the spin-$\frac{1}{2}$ case.
The unrenormalized Maxwell equation is
\begin{equation}
 {\frac{d}{d \tau}} \left({\frac {1}{\tau}}{\frac{dA}{ d \tau}} \right)
= -2e \sum_{s=1}^{4}\int [d{\bf k}](k^2_{\perp} +m^2)
\lambda_{s} {\vert f_{{\bf k}s}^{+}\vert ^2 \over \tau}\,.
\label{boost_Max_D3}
\end{equation}
Replacing $\Omega$ and $\dot\Omega$ with $\omega$ and $\dot\omega$ on
the left-hand-side of (\ref{boost_WKB_D}), we obtain the adiabatic expansion up
to second order,
\begin{equation}
\Omega_{s}^2  = \omega^2 - {\frac {1}{2 \omega^2}} \left[ \pi \ddot{\pi} +
\dot{\pi}^2 \left (1- {\frac{\pi^2}{\omega^2}}\right ) \right]
 + {\frac {3}{4}} {\frac {\pi^2 \dot{\pi}^2}{\omega^4}} +
{\frac {\dot{\pi}^2}{4 \omega^2}}
+ {\frac {\lambda_s \dot{\pi}^2 \pi }{\omega^3} }
- {\frac {\lambda_s \ddot{\pi}} {2 \omega}}\nonumber  + \cdots
\end{equation}
Using this expansion in (\ref{boost_MaxD4}) allows us to express the
integrand of (\ref{boost_Max_D3}) in the form
\begin{equation}
\sum_{s=1}^{4}(k^2_{\perp} +m^2)
(-2\lambda_{s} )){\vert f_{{\bf k}s}^{+}\vert ^2 \over \tau}
 =\frac{2\pi_{\eta}}{\tau \omega_{\bf k}} -
 \left(\frac{\ddot{\pi}_{\eta}}{2\omega^5_{\bf k}}-
\frac{5\dot{\pi}_{\eta}^2\pi_{\eta}}{4\omega^7_{\bf k}} \right )
\frac{(\omega^2_{\bf k}-\pi^2_{\eta})}{\tau} - R_{\bf k}(\tau)\,, \nonumber \\
\label{boost_R}
\end{equation}
where $R_{\bf k}(\tau)$ falls faster than $ \omega^{-3} $ and
so leads to a finite contribution to the current. Substituting (\ref{boost_R})
and using the definition (\ref{boost_omega_b}) of $\pi_{\eta}$ and its first
and second derivative into (\ref{boost_Max_D3}) yields
\begin{eqnarray}
{\frac{dE}{d \tau}} &=& {\frac{e^2}{2\tau^2}} \int [d{\bf k}] \ {\frac
{k^2_{\perp} +m^2}{\omega_{\bf k}^5}}\left \{\left(\ddot{A} -
2\frac{\dot{A}}{\tau}\right) +{\frac{5\dot{A} \pi_{\eta}^2}{\tau \omega^2_{\bf
k}}}  \right \} - e \int [d{\bf k}]R_{\bf k}(\tau)\nonumber \\
&=& -\frac{e^2}{6\pi^2}\ln\left({\Lambda\over m}\right)
{\frac{dE}{d \tau}} - e \int [d{\bf k}]R_{\bf k}(\tau). \nonumber \\
\label{boost_Max_expD}
\end{eqnarray}
where $\Lambda$ is again the cutoff in the transverse momentum integral which
has been reserved for last.

The cutoff dependent term on the right-hand side is precisely the
logarithmically divergent charge renormalization in $3 + 1$ dimensions.
Defining $\delta e^2 = (1/6\pi^2)\ln({\Lambda / m})$ as usual and
shifting this term to the left-hand side we obtain
\begin{equation}
e{\frac {dE} {d \tau}} (1+e^2 \delta e^2) =
-e^2 \int[d{\bf k}]R_{\bf k}(\tau)\,,
\label{boost_RegD}
\end{equation}
after multiplying both sides of the equation by $e$.
The renormalized charge is
\begin{equation}
e_{R}^2 = {\frac {e^2}{(1+e^2 \delta e^2)}} = Z e^2.
\label{boost_eRD}
\end{equation}
and the Ward identity assures us that $e_RE_R=eE$. Hence we obtain the
renormalized Maxwell equation
\begin{equation}
{\frac {dE_{R}} {d \tau}}  =  -e_{R} \int [d{\bf k}] R_{\bf k}(\tau),
\label{boost_jRD}
\end{equation}
where $R_{\bf k}(\tau)$ is defined by Eq. (\ref{boost_R}), and the integral is
now completely convergent.

\section{The Energy-Momentum Tensor and Effective Hydrodynamics}

In our semiclassical calculations, we follow the evolution of the
matter and electromagnetic fields. With these quantities in hand we can
calculate, in addition to the particle spectrum, other physical quantities such
as the energy density and the longitudinal and transverse pressures.
To obtain these quantities we derive the energy-momentum tensor in the
comoving frame. We shall give explicit formulae here only for the fermionic
case.

The energy-momentum tensor for QED is obtained by varying the action in
(\ref{boost_Sf}). We find
\begin{equation}
T_{\mu \nu}^{total} = {\frac {-2}{\sqrt{-g}}}{ \frac{\delta S}{\delta
g^{\mu \nu}}} = T_{\mu \nu}^{fermion} + T_{\mu \nu}^{em}
\end{equation}
with
\begin{eqnarray}
T_{\mu \nu}^{fermion}&=& \frac{i}{4} \left[ \bar {\Psi} ,\tilde{\gamma}_{(\mu}
\nabla_{\nu)} \Psi \right] - \frac{i}{4} \left[ {\nabla}_{(\mu} \bar {\Psi},
\tilde{\gamma}_{\nu)} \Psi \right]\nonumber \\
T_{\mu \nu}^{em}&=&-  \left( \frac{1}{4} g_{\mu\nu} F^{\rho \sigma} F_{\rho
\sigma} + {F_{\mu}} ^{\rho}F_{\rho \nu} \right) .
\label{c_TD}
\end{eqnarray}
In the following we shall drop the superscript on the fermion part of the
energy-momentum tensor where it causes no confusion to do so.

We are interested in calculating the diagonal terms of the matter part of the
energy-momentum tensor and in identifying them with the energy and pressure
in the comoving frame.
We begin with $T_{\tau \tau}$,
\begin{eqnarray}
\langle 0| T_{\tau \tau} |0 \rangle
&=& \ {\frac{i}{\tau}}\sum_{s=1}^{2}\int [d{\bf k}]\, (k^2_{\perp}
+m^2)\left[ f^{\ast -}_{{\bf k}s} \stackrel{\leftrightarrow}{{d \over d \tau}}
f^{-}_{{\bf k}s} - f^{\ast +}_{{\bf k}s}\stackrel{\leftrightarrow}{{d \over d
\tau}} f^{+}_{{\bf k}s}\right] \nonumber \\
&=&  -{\frac{i}{\tau}}\sum_{s=1}^{2}\int [d{\bf k}]\,\left[2(k^2_{\perp}
+m^2) f^{\ast +}_{{\bf k}s} \stackrel{\leftrightarrow}{{d \over d \tau}}
f^{+}_{{\bf k}s}+\frac{\lambda_s\pi_{\eta}} {\tau}\right].
\label{c_Ttau1}
\end{eqnarray}
In the latter form only the positive frequency mode functions appear which is
most useful for the adiabatic expansion below.
Averaging over  $s=1,2$ and $s=3,4$ we may also write (\ref{c_Ttau1}) in the
form
\begin{eqnarray}
\langle 0| T_{\tau \tau}|0 \rangle
  = - 2\sum_{s=1}^{4}\int [d{\bf k}](k_{\perp}^2+m^2){\Omega_{{\bf k}s}\over
\tau} \vert f_{{\bf k}s}^{+}\vert ^2 .
\label{c_Ttau2}
\end{eqnarray}
This expression contains quartic and quadratic divergences in $3 + 1$
dimensions present even in the complete absence of fields (the vacuum energy or
cosmological constant terms) and a logarithmic divergence which is related to
the charge renormalization of the last section.
To isolate these divergent terms we express the integrand of $T_{\tau\tau}$ as
the sum of its second order adiabatic expansion and a remainder term,
\begin{equation}
- 2\sum_{s=1}^{4}(k_{\perp}^2+m^2){\Omega_{{\bf k}s}\over \tau} \vert f_{{\bf
k}s}^{+}\vert ^2 = -\frac{2\omega}{\tau}
+ (k^2_{\perp} +m^2) {\frac {(\pi_{\eta} +e \dot{A})^2}
{4 \omega^5 \tau^3}} + R_{\tau \tau}({\bf k}),
\label{c_TtauReg}
\end{equation}
where $ R_{\tau \tau}({\bf k})$ falls off faster than
$\omega^{-3}$ so that the integral over $ R_{\tau \tau}({\bf k})$ is finite.

The first term in (\ref{c_TtauReg}) gives rise to a quartic divergence in $3$
space dimensions (or a quadratic divergence in $1$ space dimension) independent
of the electric field and must be subtracted. The $\pi_{\eta}^2$ term in
(\ref{c_TtauReg}) gives rise to a quadratic divergence in $3$ dimensions which
must be likewise subtracted. However, in $1$ space dimension this term yields a
{\it finite} contribution to $\langle 0| T_{\tau \tau}|0 \rangle$ which must be
retained, since it is $\tau$ dependent, and cannot be absorbed into a
cosmological constant counterterm. Strictly speaking, subtracting this term in
$3$ dimensions can only be justified by using a coordinate invariant
regularization scheme for formally divergent quantities, such as dimensional
regularization, where quartic and quadratic divergences are automatically
excised. Only in such a scheme can one be certain that the divergence can be
absorbed into a counterterm of the generally coordinate invariant lagrangian
in (\ref{boost_Sf}). The net result is that this term must be handled somewhat
differently in the $3$ and $1$ dimensional cases.

The term which is linear in $\dot{A}$ vanishes when integrated
symmetrically. The term in (\ref{c_TtauReg}) proportional to $e^2 {\dot{A}^2}$
is logarithmically divergent in $3$ dimensions, and finite in $1$ dimension. In
fact it is precisely
\begin{equation}
\delta e^2 \frac { \dot{A}^2}{2 \tau ^2}\,,
\label{c_Tdiv}
\end{equation}
where $\delta e^2$ is given by (\ref{chgeren}) in $3$ dimensions and by
$(12\pi m^2)^{-1}$ in $1$ dimension. In either case, it can be absorbed into a
renormalization of the electric energy term of the stress tensor,
\begin{equation}
T_{\tau \tau}^{em}=\frac {\dot{A}^2}{2 \tau ^2}\ .
\end{equation}
by charge renormalization. When added to the electromagnetic term it gives
\begin{equation}
(1 + e^2\delta e^2 )\frac { \dot{A}^2}{2 \tau ^2}=
Z^{-1}\frac { \dot{A}^2}{2 \tau ^2}= {\frac{E_{R}^2}{2}}.
\label{c_Trem}
\end{equation}
Thus the explicitly finite, renormalized $\langle T_{\tau \tau}\rangle$ for the
combined matter and electromagnetic system is
\begin{eqnarray}
\langle T_{\tau \tau} \rangle = {\frac{E_R^2}{2}}+
\int [d{\bf k}] R_{\tau\tau}({\bf k})\,,
\label{energytt}
\end{eqnarray}
in three dimensions where $R_{\tau\tau}({\bf k})$ is defined in
{}~(\ref{c_TtauReg}). In one dimension the finite $\pi_{\eta}^2$ term that must
be retained in the adiabatic expansion of (\ref{c_TtauReg}) gives rise to an
additional $(12\pi\tau^2)^{-1}$ on the right side of (\ref{energytt}).

Turning now to $T_{\eta \eta}$, the matter part is given by
\begin{equation}
\langle 0| T_{\eta \eta}|0\rangle  =  2\tau \sum_{s=1}^{4}\int [d{\bf
k}](k^2_{\perp} +m^2)  \lambda_s \pi_{\eta}  \vert
f_{{\bf k}s}^{+}\vert ^2 .
\label{c_Teta}
\end{equation}
The adiabatic expansion of the integrand gives in this case
\begin{eqnarray}
2\tau \sum_{s=1}^{4}(k^2_{\perp} +m^2)  \lambda_s \pi_{\eta}  \vert
f_{{\bf k}s}^{+}\vert ^2&=& -\frac{2\pi^2_{\eta}\tau}{ \omega_{\bf k}}+
 \left[\frac{\ddot{\pi}_{\eta}}{2\omega^5_{\bf k}}-
\frac{5\dot{\pi}_{\eta}^2\pi_{\eta}}{4\omega^7_{\bf k}} \right ]
{\pi_{\eta}\tau(\omega^2_{\bf k}-\pi^2_{\eta})} \nonumber \\
&&\quad\hbox{}+\pi_{\eta}\tau^2 R_{\bf k}(\tau) \,.
\label{c_Reta}
\end{eqnarray}
Inserting this into (\ref{c_Teta}) we find again that the quadratic and quartic
divergences in $3$ dimensions are independent of the electric field
and that the terms proportional to $\dot{A}$ vanish, whereas the term
proportional to $ \dot{A}^2 $ again renormalizes the electromagnetic
contribution to the energy-momentum tensor,
\begin{equation}
T_{\eta \eta}^{em}=-\frac {\dot{A}^2}{2}\ .
\end{equation}
Therefore the fully renormalized total $T_{\eta \eta}$ in $3$ dimensions is
\begin{equation}
\langle T_{\eta \eta} \rangle=-{\frac12} E_{R}^2 \tau^2 + \tau^2 \int [d{\bf
k}]\,\pi_{\eta}  R_{\bf k}(\tau)\,.
\label{c_TReta}
\end{equation}
As for $\langle T_{\tau \tau} \rangle$ there is a finite additional term in $1$
dimension that must be added to this expression, which in this case is equal to
a constant, $(12\pi)^{-1}$.

For the transverse components of the matter stress tensor in $3$ dimensions we
have
\begin{eqnarray}
\langle T_{xx} \rangle &=& \langle T_{yy} \rangle  \nonumber \\
&=&
-{\frac{1}{2\tau}}\sum_{s=1}^{4}\int [d{\bf k}]k^2_{\perp}
\left [ \left(i f^{\ast +}_{{\bf k}s}
\stackrel{\leftrightarrow}{d\over d\tau}f^{+}_{{\bf k}s} \right)
+ \lambda_s \pi_{\eta}  \vert f_{{\bf k}s}^{+}\vert ^2  \right ].
\label{c_Txx}
\end{eqnarray}
In a precisely analogous manner one develops the adiabatic expansion of the
integrand, isolates the divergences and the logarithmically divergent charge
renormalization which combines with the electromagnetic stress,
\begin{equation}
T_{xx}^{em}=T_{yy}^{em}=\frac {\dot{A}^2}{2 \tau ^2}\ .
\end{equation}
and arrives at the renormalized form for the total $\langle T_{xx} \rangle =
\langle T_{yy} \rangle$
can be written as
\begin{equation}
\langle T_{xx} \rangle = \frac12 \int [d{\bf k}]
 {\frac{k^2_{\perp}}  {(k^2_{\perp}
+m^2)}} [ R_{\tau \tau} -  \pi_{\eta}  R_{\bf k} ]
+{\frac {1}{2}}E_{R}^2\,
\label{c_TRxx}
\end{equation}
where $R_{\tau \tau}$ and $R_{\bf k}$ have been defined previously. This result
for the transverse components may be obtained by consideration of the trace of
the energy momentum tensor $T^{\mu}_{\mu}$. From Eqs.~(\ref{energytt}),
(\ref{c_TReta}), and (\ref{c_TRxx}) we note that this trace vanishes as $m
\rightarrow 0$.

Both the unrenormalized and renormalized total energy-momentum
tensors are covariantly conserved, so that we have
\begin{eqnarray}
T^{\mu\nu}_{\quad ;\mu}&=& (T_{\quad ;\mu}^{\mu\nu})_{matter}
+ (T_{\quad ;\mu}^{\mu\nu})_{em}=0\,, \nonumber \\
 (T_{\quad ;\mu}^{\mu\nu})_{em}&=&-F^{\nu}_{\, \mu}j^{\mu} .
\label{c_nablaT}
\end{eqnarray}
In the boost invariant proper time coordinates one may verify that this
equation takes the form,
\begin{equation}
\partial_{\tau} T_{\tau \tau} +\frac {T_{\tau\tau}}{\tau} +
\frac {T_{\eta \eta}}{\tau^3} = F_{\eta \tau} j^{\eta}\,.
\label{c_Tconserv}
\end{equation}
If we follow the standard practice and define the energy density and transverse
 and longitudinal pressures via,
\begin{equation}
T_{\mu \nu} = {\rm diag} (\epsilon, p_{\perp},p_{\perp}, p_{\|}\tau^2),
\label{stress}
\end{equation}
then the energy conservation equation takes the form
\begin{equation}
{\frac{d(\epsilon \tau)}{d\tau}} + p_{\|} = E j_{\eta}
\label{Tconsv}
\end{equation}

In most hydrodynamical models of particle production, one usually assumes that
thermal equilibrium sets in and that there is an equation of state
$p_{\|} = p_{\|}(\epsilon)$.  For boost-invariant kinematics  $v=z/t$, all the
collective variables are functions only of $\tau$ and therefore $p_{\|}$ is
implicitly a function of $\epsilon$. In our field-theory model
in the semiclassical limit there is no real scattering of partons and
thus one does not have charged particles in equilibrium or a true equation of
state. Hence the transverse and longitudinal pressures are different. In the
next order in $1/N$ there is parton-parton scattering and it remains to be seen
whether  thermal equilibrium and isotropy of the pressures will emerge
dynamically.  Nevertheless, even in this order in $1/N$ one can define an
effective hydrodynamics using the expectation value of the field theory's
stress
tensor (\ref{stress}).  Formally introducing the auxiliary quantities
``temperature" and ``entropy density" in a suggestive way in analogy with
thermodynamics via,
\begin{equation}
\epsilon +  p_{\|} =Ts\ ; \qquad d\epsilon = T ds
\label{entropy}
\end{equation}
we find that the entropy density obeys,
\begin{equation}
{\frac{d(s \tau)}{d\tau}} = {\frac { E j_{\eta}} {T}}
\label{dconsv1}
\end{equation}
Notice that when the electric field goes to zero
$s \tau$ becomes constant. If $ p_{\|}$ also goes
to zero with $\tau$ faster than $1/\tau$ ,  then $ \epsilon \tau$ is
constant when the electric field goes to zero.

In standard phenomenological models of particle production such as Landau's
hydrodynamical model, one usually assumes that the hydrodynamics describes
an isotropic perfect fluid whose energy momentum tensor in a comoving frame
has the form (\ref{stress}) with $p_{\|}= p_{\perp}$. Then one component of the
energy conservation equation becomes,
\begin{equation}
{\frac{d(s \tau)}{d\tau}} = 0
\label{dconsv2}
\end{equation}
which is of the same form as (\ref{dconsv1}) in the absence of electric field,
with the difference that the {\it isotropic} pressure enters into the
definitions of the entropy density and temperature of the fluid in
(\ref{entropy}). From these definitions one can also calculate directly the
entropy density in the comoving frame as a function of $\tau$ by
\begin{equation}
s(\tau) = \exp \left \{ \int_0^{\tau} {1 \over \epsilon + p} {d\epsilon \over d
\tau} \ d \tau \right \}
\label{ent1}
\end{equation}
Since we follow the microscopic degrees of freedom, we can also construct the
Boltzmann entropy function in terms of the single particle distribution
function in comoving phase space. This is done in Appendix A.

Let us compare the energy spectra of the isotropic hydrodynamics
with the results of our field theoretic approach. One quantity we wish to
determine is the amount of lab frame energy in a bin of fluid rapidity,
$dE_{lab} / d\eta$.  For the isotropic hydrodynamics, in the lab frame one can
write the energy momentum tensor in the form,
\begin{equation}
T^{ab} = (\epsilon + p) u^{a}u^{b}
+p\eta^{ab}
\end{equation}
where $u^{t} = \cosh \eta$ and $u^{z} = \sinh \eta$.
Then calculating $dE_{lab} / d\eta$ on a surface of constant proper time $\tau$
we obtain,
\begin{eqnarray}
{dE_{lab} \over d\eta} & = & \int T^{t a} {d\sigma_{a} \over d\eta} \nonumber
\\ d\sigma_{a} & = & A_{\perp}(dz, -dt) = A_{\perp} \tau_{f}(\cosh\eta,
-\sinh\eta)  \nonumber \\
{dE_{lab} \over d\eta} & = & A_{\perp}\epsilon(\tau_{f}) \tau_{f}\cosh \eta \ ,
\label{Espectrum}
\end{eqnarray}
where $A_{\perp}$ is a transverse size which in a flux tube model would
be the transverse area of the chromoelectric flux tube.

We show now that our microscopic hydrodynamics gives an {\it identical} result,
without any assumptions about thermal equilibrium. In fact, transforming the
result of our field theory calculation (\ref{stress}) to the lab frame,
\begin{eqnarray}
T^{ab}
&=&\pmatrix{ \epsilon \cosh^2 \eta + p_{\|} \sinh^2 \eta
 & 0&0&(\epsilon +p_{\|}) \cosh \eta \sinh \eta\cr
0&p_{\perp}&0&0\cr
0&0&p_{\perp}&0 \cr
(\epsilon +p_{\|}) \cosh \eta \sinh \eta
&0&0&\epsilon \sinh^2 \eta + p_{\|} \cosh^2 \eta \cr} \nonumber \\
\label{boost_Tab}
\end{eqnarray}
and recalculating ${dE_{lab} / d\eta}$ again gives (\ref{Espectrum}), where
$\epsilon (\tau)$ is now explicitly calculable from the modes of the field
theory.

In hydrodynamic models one assumes that hadronization does not effect the
collective motion. If all the particles that are produced after hadronization
are pions then the number of particles in a bin of rapidity should be
just the energy in a bin of rapidity divided by the energy of a single pion
namely,
\begin{eqnarray}
{dN \over d\eta} &= &{ 1 \over m_{\pi} \cosh\eta} {dE_{lab} \over d\eta}
\nonumber \\ & = & A_{\perp} {\epsilon(\tau_{f})\tau_{f} \over m_{\pi}}
\label{pionnum}
\end{eqnarray}
To see if this formula is working in our parton domain we can instead use the
mass of a parton in place of $m_{\pi}$ and check directly whether the spectrum
of partons given by
\begin{equation}
{dN \over d\eta} = {\epsilon(\tau_{f}) \tau_{f}\over m}
\label{partnum}
\end{equation}
agrees with explicit calculation of particle number in the field theory, as
given in Appendix A.

In order for the formula (\ref{partnum}) to be independent of $\tau_f$ we
require that the electric field become vanishingly small and that
the pressure go
to zero faster than $\frac {1}{\tau}$ at large $\tau$.  Indeed we will find
this is approximately true in the numerical simulations, the results of which
we will present in the next section.

\section{Numerical results in (1+1) dimensions}

In this section we present the results of solving the back-reaction problem
in two dimensions (proper time~$\tau$ and fluid rapidity~$\eta$), and compare
the results to a phenomenological Boltzmann-Vlasov model. In previous
calculations using kinetic equations in flux tube models
\cite{BC86,BC88,Kajantie} it was assumed that the Schwinger source term (WKB
formula) can be used by taking the electric field, hitherto constant, to be a
function of proper time. However, in the Schwinger derivation the time
parameter (which is not boost invariant) plays an implicit role. Therefore, it
is not clear {\em a priori\/} if such a source term in the kinetic equations
represents the correct rate of particle production. From the experience
obtained in the spatially homogeneous case, we believe that if we know the
correct source term, a phenomenological Boltzmann-Vlasov approach should agree
with the semiclassical QED calculation.

The phenomenological Boltzmann-Vlasov equation in $3+1$ dimensions
can be written covariantly as
\begin{equation}
{D f \over D \tau} \equiv p^{\mu}\frac{\partial f}{\partial q^{\mu}}
-e p^{\mu}F_{\mu\nu}\frac{\partial f}{\partial p_{\nu}} = p_0 g^{00} \frac{dN}
{dq^0 d^3{\bf q} d^3 {\bf p}}\, ,
\label{boostBV}
\end{equation}
We shall write the transport equation in the comoving coordinates
and their conjugate momenta,
\begin{eqnarray}
q^{\mu}=(\tau,x,y,\eta),\quad
p_{\mu}=(p_{\tau},p_x,p_y,p_{\eta})\,.
\label{boost_qp}
\end{eqnarray}
In order to write the invariant source term in these coordinates,
we begin with the WKB formula, which is
\begin{eqnarray}
\frac{dN}{[(-g)^{1/2}dq^0 d^3{\bf q}] d^2 {\bf {p_{\perp}}}}
&=& \pm [1\pm 2f({\bf{p}},\tau)] e \vert E(\tau)\vert\ \nonumber \\
&&\times \ln {\left [1 \pm \exp{\left (-\frac{\pi( m^2+{\bf p}^2_{\perp})}
{e\vert E(\tau)\vert} \right )}\right ]},\nonumber\\
\label{boost_source3}
\end{eqnarray}
if the (constant) electric field is in the $z$ direction. The $\pm$ refers to
the cases of charged bosons or fermions respectively.
Our model for the spatially homogeneous case consisted of applying this
formula even for a time-dependent electric field. Here we will allow the
electric field to be a function of the proper time, writing
\begin{equation}
\sqrt{F^{\mu\nu}F_{\mu\nu}}=\vert E(\tau)\vert\,.
\label{boost_eE}
\end{equation}

In the spatially homogeneous case we assumed that particles are produced at
rest, multiplying the WKB formula by $\delta (p_z)$\@.
This longitudinal momentum dependence violates the Lorentz-boost symmetry.
Here we assume \cite{BC86,BC88,Kajantie} that the $p_{\eta}$ distribution is
$\delta (p_{\eta})$, which is boost-invariant according to
(\ref{boost_ptaueta}).
Assuming boost-invariant initial conditions for $f$, invariance
of the Boltzmann-Vlasov assures that the distribution function
is a function only of the boost invariant variables ($\tau,\eta-y$)
or ($\tau,p_\eta$). The kinetic equation reduces to
\begin{eqnarray}
\frac{\partial f}{\partial
\tau}+eF_{\eta\tau}(\tau)\frac{\partial f}
{\partial p_{\eta}}
&=&\pm [1\pm 2f({\bf{p}},\tau)] e\tau \vert E(\tau)\vert \nonumber \\
&& \times \ln {\left [1 \pm \exp{\left (-\frac{\pi
(m^2+{\bf{p}}^2_{\perp})}{e\vert E(\tau)
 \vert }\right )}\right ]}\delta (p_{\eta}).\nonumber \\
\label{boost_BV}
\end{eqnarray}

Turning now to the Maxwell equation, we have from
(\ref{boost_Maxb}) that
\begin{eqnarray}
-\tau\frac{dE}{d\tau}=j_{\eta}=j^{cond}_{\eta}+j^{pol}_{\eta}\,,
\label{boost_BVmax}
\end{eqnarray}
where $j^{cond}$ is the conduction current and
$j^{pol}_\mu$ is the polarization current due to pair creation
\cite{Gatoff,PRL,PRD}.
The invariant phase-space in the comoving coordinates is
\begin{eqnarray}
{\frac {1} {(-g)^{1/2}p_0 g^{00}}}{\frac{d^3 {\bf p}}{(2\pi)^3 }} =
\frac{d{\bf{p}}_{\perp}dp_{\eta}}{(2\pi)^3 \tau p_{\tau}}\,.
\label{boost_phase}
\end{eqnarray}
Thus in (1+1) dimensions we have
\begin{eqnarray}
j^{cond}_{\eta}&=&2e\int \frac{dp_{\eta}}{2\pi \tau p_{\tau}}p_{\eta}
f(p_{\eta},\tau) \nonumber \\
j^{pol}_{\eta}&=&\frac{2}{F^{\tau\eta}}\int \frac{dp_{\eta}}
{2\pi \tau p_{\tau}} p^{\tau}{D f \over D \tau} \nonumber\\
&=& \pm [1\pm 2f(p_{\eta}=0,\tau)]\frac{m e\tau}{\pi}
{\rm sign}[E(\tau)] \ln {\left [1\pm \exp{\left (-\frac{\pi m^2}
{\vert eE(\tau) \vert} \right )}\right ]}.\nonumber \\
\label{boost_jpol}
\end{eqnarray}

Assuming that at $\tau=\tau_i$ there are no particles, the solution of
(\ref{boost_BV}) along the characteristic curves
\begin{eqnarray}
\frac{dp_{\eta}}{d\tau}=eF_{\eta\tau}(\tau)
\label{boost_char}
\end{eqnarray}
is
\begin{eqnarray}
f(p_{\eta},\tau)&=&\pm \int_{\tau_i}^{\tau} d\tau'\, [1\pm
2f(p_{\eta}=0,\tau')]
\, e\tau' \vert E(\tau')\vert
 \\
&&\qquad\hbox{}\times \ln {\left [1 \pm \exp{\left (-\frac{\pi m^2}{e\vert
E(\tau') \vert }
\right )}\right ]}\delta (p_{\eta}-eA_{\eta}(\tau')+eA_{\eta}(\tau)).\nonumber
\label{boost_fBV}
\end{eqnarray}
Thus the system (\ref{boost_BV})--(\ref{boost_BVmax}) reduces to
\@\footnote{
A related derivation (but without the generality and fully covariance of
the present one)
for this model in terms of different variables
can be found in \cite{BC86,BC88,Kajantie}.}
\begin{eqnarray}
-{\tau }\frac{d {E}}{d{\tau}}&=&
\pm \frac{e^2}{\pi m^2}\int_{{\tau}_i}^{{\tau}}\,d{\tau'}
 [1\pm 2f(p_{\eta}=0,\tau')]\,
\frac{A({\tau'})-A({\tau})}
{\sqrt{\left[A({\tau'})-A
({\tau})\right]^2+{\tau}^2}}{\tau'}\nonumber\\
&&\qquad\qquad\hbox{}\times\left| {E}({\tau'})\right|
\ln \left[1 \pm \exp \left(-\frac {\pi}{\left|{E}({\tau'})
\right|}\right)\right]\nonumber\\
&&\pm [1\pm 2f(p_{\eta}=0,\tau)]\,\frac{{\tau} e^2}{\pi m^2}\,{\rm sign}
\left({E}({\tau})\right)
\ln \left[1+\exp\left(-\frac {\pi}{\left|{E}({\tau})\right|}
\right)\right] .\nonumber \\
\label{boost_jBVtau}
\end{eqnarray}
In the above expression and in the following we introduce the dimensionless
variables.
\begin{equation}
m\tau \rightarrow {\tau}\,, \qquad
eA \rightarrow {A}\,,\qquad
{\frac{eE}{m^2} }\rightarrow {E} \,,\qquad
{ej_{\eta} \over m^2} \rightarrow  j_\eta \,,
\label{boost_dimensionless}
\end{equation}

For scalar particles we carry out the calculations in terms of the conformal
proper time $u$ (see Appendix C). This gives us more control over the physics
at very early times, allowing us to choose a well-behaved initial adiabatic
vacuum, corresponding to initial conditions
\begin{eqnarray}
W_{\bf k} (u_0) &=& w_{\bf k} (u_0) ,\nonumber \\
\dot{W}_{\bf k} (u_0) &=& \dot{w}_{\bf k} (u_0)
\end{eqnarray}
It is worth mentioning that the adiabatic vacuum in terms
of $u$ is not identical to the adiabatic vacuum in terms of $\tau$;
they are related by a Bogolyubov transformation.
The variable $u$ is regular and improves numerical
stability near the singular point $\tau=0$.

For an initial adiabatic vacuum state the renormalized Maxwell equation is
given by
\begin{eqnarray}
-\frac{d E}{du}=
\frac{e^2_R/m^2}{1-e_R^2\delta e^2}
\int_{-\infty}^{\infty}\frac{dk_{\eta}}{2\pi}
(k_{\eta}-A)\left[ \frac{1}{W_{k_{\eta}}(u)}-\frac{1}{w_{k_{\eta}}(u)}
\right],
\label{boost_1D_jR}
\end{eqnarray}
where $\delta e^2 = (12\pi m^2)^{-1}$. Equations (\ref{boost_1D_jR}) and
(\ref{boost_uWKB}) define the numerical problem for the boson case.
In solving (\ref{boost_1D_jR}) and (\ref{boost_uWKB}) we discretize the
momentum variable in a box with periodic boundary conditions,
$k_{\eta} \rightarrow \pm 2 \pi n / L$ where $L=500$ and $n$ ranges
from $1$ to $3\times 10^4$. The time step in $u$ was taken to be
$5\times 10^{-4}$.

To compare the Boltzmann-Vlasov phenomenological model to the above
semiclassical system, Eq.~(\ref{boost_jBVtau}) is written
in terms of the conformal proper time variable $u$ and becomes
\begin{eqnarray}
-\frac{d {E}}{du}&=&
\pm \frac{e^2}{\pi m^2}\int_{u_i}^{u}\,d{u'}
[1\pm 2f(p_{\eta}=0,u')]\frac{A({u'})-A(u)}
{\sqrt{\left[A({u'})-A
(u)\right]^2+e^{2u}}}e^{2{u'}}\nonumber\\
&&\qquad\qquad\hbox{}\times\left| {E}({u'})\right|
\ln \left[1\pm\exp \left(-\frac {\pi}{\left|{E}({u'})
\right|}\right)\right]\nonumber\\
&&\pm [1\pm 2f(p_{\eta}=0,u)]
\frac{e^{u} e^2}{\pi m^2}\,{\rm
sign}\left({E}(u)\right)
\ln \left[1\pm\exp\left(-\frac {\pi}{\left|{E}(u)\right|}
\right)\right] ,\nonumber \\
\label{boost_jBVu}
\end{eqnarray}
where ${dA}/{du}=-e^{2u}{E}$\@.

In the fermion problem we perform the simulations in terms of $\tau$,
and in this  case the semiclassical problem is defined by Equation
(\ref{boost_modef_D}) and by the Maxwell equation
\begin{equation}
\tau \frac{dE(\tau)}{d\tau} = -\frac{2(e^2_R/m^2)}{1-e_R^2 \delta e^2}
\sum_{s=1}^{2}\int
 \frac{dk_{\eta}}{2\pi}
\lambda_{s}\vert f_{{\bf k}s}^{+}\vert ^2 .
\label{boost_MaxD1b}
\end{equation}
In this $1+1$ dimensions problem  $\delta e^2 = (6\pi m^2)^{-1}$,
$\lambda_1=1$, and $\lambda_2=-1$. The time step in $\tau$ was taken to be
$0.0005$ with the momentum grid the same as for the scalar problem.

Figs. ~1 and ~2 summarize the results of the numerical simulations of charged
scalar particles in $1 + 1$ dimensions.  In Fig.~1 we show the time evolution
of $A(u)$, $ {E}(u)$ and~$ {j}_{\eta}(u)$ for the case of ${E}(u=-2)=4$
and~$e^2/m^2=1$\@. We see that in the first oscillation the electric field
decays quite strongly, in contrast with the spatially homogeneous case
\cite{PRL}. In the latter the degradation of the electric field results from
particle production only. In this inhomogeneous problem our system expands, and
hence the initial electromagnetic energy density is reduced due to the particle
production and this expansion. This degradation due to the expansion can be
inferred by solving a classical system of particles and antiparticles that
interact with a proper-time dependent electric field without a particle
production source term. We also note that the larger the initial field, the
smaller the period of oscillations.

In Fig.~2 we compare the time evolution of the phenomenological model (dashed
curve) with the result of the semiclassical calculation (solid curve). Initial
conditions were fixed at $u=-2$ [Figs.~2(a)--2(b)] and at $u=0$
[Figs.~2(c)--2(d)]. We see that there is good agreement between the
semiclassical solution and the Boltzmann-Vlasov model. This agreement also
holds for different values of the initial electric field and for different
coupling constants. The Bose enhancement increases the frequency, and hence a
better agreement is achieved, as expected. We conclude that the WKB formula
with Bose enhancement is a suitable source term for the boost-invariant
problem.
It is worth mentioning that at very early times of the evolution
(before $\tau=1$) the particle production is negligible and
the electric field falls very slowly, as can bee seen in Figs.~2(c)--2(d).

In Figs.~3 through ~9 we present the numerical results for fermions in $1 + 1$
dimensions. In Fig.~3 we compare the time evolution of the Boltzmann-Vlasov
equation (dashed curve) with the results of the semiclassical calculation
(solid curve), where the initial conditions were fixed at $\tau=1$. All
succeeding figures refer to these same initial conditions.
In Fig.~4 we present the time evolution for $\tau \epsilon $ where
$\epsilon=T_{\tau\tau}$. We see that at large $\tau$ this quantity oscillates
around a fixed value. In Fig.~5 we show the time evolution
of $p/{\epsilon}$ where $p\tau^2=T_{\eta\eta}$. In this lowest order
calculation there is no true dissipation and no particular equation of state
emerges from the time evolution, although there is some indication that $p$
approaches zero faster than $\epsilon$. In Fig.~6 we present the
evolution of the particle density $dN/d\eta$. After a short time (of order
$\tau = 15$) the particle density reaches a plateau which doesn't change much
in the subsequent evolution. This is consistent with the fact that the
Schwinger particle creation mechanism turns off rapidly as the electric
field decreases. In Fig.~7 we present the time evolution of $\tau \epsilon
/(dN/d\eta)$. One can see that at large $\tau$ there is some indication that
this ratio approaches the value of the mass (we choose $m=1$), which agrees
with the prediction of the hydrodynamic model discussed in Section 4.
[see Eq. (\ref{partnum})]. This lends support to the idea that the pion
spectrum can be calculated using (\ref{pionnum}). Defining the Boltzmann
entropy by (\ref{ent2}) of Appendix A, we plot $\tau s$ as a function of $\tau$
in Fig.~8. Notice that this quantity is roughly constant after $\tau \approx
20$, by which time particle production has nearly ceased. [Compare Fig.~6.]
This agrees with the result expected from the hydrodynamic point of view, eq.
(\ref{dconsv2}). Finally, in Fig.~9, we plot the effective ``temperature,"
defined by the hydrodynamic relation (\ref{entropy}), but using the Boltzmann
entropy of Fig.~8.

In conclusion, the present results using boost-invariant
coordinates fall into line with previous studies \cite{PRL,PRD}
of boson and fermion pair production from an electric field with
back-reaction. The renormalized field-theory calculation is
tractable, yielding oscillatory behavior for a relativistic
plasma which can also be well described by means of a classical
transport equation with a source term derived from the Schwinger
mechanism modified by Bose enhancement or Pauli blocking.  For
the boost-invariant case the electric field decays much more
rapidly than for cartesian coordinates, where the sole decay
mechanism is transfer of energy to the produced pairs.  The
ability to use the transport-equation approximation for
boost-invariant pair production justifies in part the use that
has been made of this method of description in past studies of
the production of the quark-gluon plasma, and opens the way for
further applications in the future.

\newpage

\newcounter{mysec}
\newcommand{\myappendix}{\appendix
\setcounter{mysec}{0}
 \renewcommand{\themysec}{\Alph{mysec}}}
\newcommand{\myappsection}[1]{
 \setcounter{equation}{0}
  \addtocounter{mysec}{1}
\section*{\themysec\ \ #1}}
\renewcommand{\theequation}{\themysec.\arabic{equation}}

\myappendix

\myappsection{The particle spectrum of fermions}

We present here the formulae for direct calculation of the fermion particle
spectrum in the adiabatic method.  For further discussion see \cite{PRD,Revs}.

During particle production particle number is not
conserved nor even uniquely defined. In terms of the adiabatic expansion of
Sections 3 and 4, however, a natural definition of an interpolating
particle-number operator suggests itself. One may simply expand the field in
terms of the {\it time-dependent} creation and annihilation operators of the
lowest order adiabatic vacuum,
\begin{equation}
\Psi (x) = \int [d{\bf k}] \sum_{s}[a_{s}({\bf k};\tau)
y^{+}_{{\bf k}s}(\tau) e^{ i {\bf{k}} \cdot {\bf x}}
+c_{s}^{\dagger}({\bf{-k}};\tau) y^{-}_{{\bf{-k}}s}(\tau)
 e^{ - i {\bf{k}} \cdot {\bf x}}  ].
\label{boost_fieadb}
\end{equation}
where
\begin{equation}
y^{\pm}_{{\bf k}s} = \left[-\gamma^{0}\left( {d \over d\tau}
+\frac{1}{2\tau}\right)
- i \gamma_{\bf{{\perp}}}\cdot {\bf{k_{\perp}}}
-i \gamma^{3} \pi_{\eta}+ m\right] \chi_{s} {g^{\pm}_{{\bf k}s} \over {\sqrt
\tau}}
\, , \label{boost_psi_ga}
\end{equation}
analogously to eq. (\ref{boost_psi_g}) of the text, but in which the
exact mode functions $f^{\pm}_{{\bf k}s}$ obeying (\ref{boost_modef_D}) are
replaced by lowest order adiabatic mode functions $g^{\pm}_{{\bf k}s}$. The
positive frequency adiabatic mode function is given explicitly by  substituting
$\omega_{\bf k}(\tau)$ for $\Omega_{\bf k}(\tau)$ in the expressions
(\ref{boost_ansatz_D}) and (\ref{boost_MaxD4}) for $f^{\pm}_{{\bf k}s}$ in the
text.

The adiabatic basis functions and the corresponding Fock space particle
annihilation and creation operators, $a_{s}({\bf k};\tau)$ and
$c_{s}^{\dagger}({\bf{-k}};\tau)$ defined in this way are related to those
defined in (\ref{boost_fieldD}) by a time-dependent Bogoliubov transformation.
This transformation is easily found by using the Dirac inner product,
\begin{equation}
(u,v) \equiv \int d\Sigma^{\mu} \bar {u} \tilde {\gamma}_{\mu} v =\int d^d {\bf
x} \sqrt{-g} \ u^{\dagger} v\, .
\label{innr}
\end{equation}
Indeed by substituting the two expansions of the field operator in terms of the
two different bases (\ref{boost_fieldD}) and (\ref{boost_fieadb}) we find
\begin{equation}
a_{s}({\bf k};\tau) = (y^{+}_{{\bf k}s}(\tau) e^{ i {\bf{k}} \cdot {\bf x}},
\Psi) = \alpha ({\bf k}s ; \tau) b_{s}({\bf k}) + \beta^{\ast} ({\bf k}s ;
\tau) d^{\dagger}_{s}(-{\bf k})\, ,
\label{Bogl}
\end{equation}
with
\begin{equation}
\alpha ({\bf k}s ; \tau) = 2 \Bigl\{ {d  g^{+ \ast}_{{\bf k}s}\over d \tau}
{d  f^{+}_{{\bf k}s}\over d \tau} + i \lambda_s \pi_{\eta} \Bigl( g^{+
\ast}_{{\bf k}s} {\stackrel {\leftrightarrow}{d \over d \tau}} f^{+}_{{\bf k}s}
\Bigr) + \omega^2_{\bf k} g^{+ {\ast}}_{{\bf k}s} f^{+}_{{\bf k}s} \Bigr\}\, .
\label{Bogcoefa}
\end{equation}
Squaring this expression and using (\ref{boost_ansatz_D}) and its analog for
the adiabatic mode function $g^{+}_{{\bf k}s}$, we arrive at
\begin{eqnarray}
\vert\beta ({\bf k}s ; \tau)\vert^2 &=& 1 - \vert\alpha ({\bf k}s ;
\tau)\vert^2  \nonumber \\
&=& 4 \vert f_{{\bf k}s}^{+}\vert ^2 \vert g_{{\bf k}s}^{+}\vert ^2
(\omega_{\bf k}^2 - \pi_{\eta}^2) \nonumber \\
&&\qquad\hbox{}\times  \Bigl\{ (\Omega_{{\bf k}s} - \omega_{{\bf k}})^2 +
\Bigl[ {{(\dot{\Omega}_{{\bf k}s}} + \lambda_s {\dot{\pi}_{\eta}}) \over 2
\Omega_{{\bf k}s}} - {{(\dot{\omega}_{{\bf k}}} + \lambda_s {\dot{\pi}_{\eta}})
\over 2 \omega_{{\bf k}}} \Bigr]^2 \Bigr\}\, . \nonumber \\
\label{Bogcoefb}
\end{eqnarray}
The expectation value of the number operator with respect to the adiabatic Fock
space operators in (\ref{boost_fieadb}) is then simply the sum over $s= 1, 2$
or $s= 3, 4$ of
\begin{eqnarray}
N({\bf k}s,\tau) &=& N_{+}({\bf k}s) | \alpha({\bf k}s; \tau) |^{ 2}
+(1-N_{-}({\bf k}s))|\beta({\bf k}s ;\tau)|^{2}\nonumber\\
&+& 2 {\rm Re}\{\alpha({\bf k}s; \tau)\beta({\bf k}s ;\tau) F({\bf k}s)\} .
\label{s_A12}
\end{eqnarray}
For initial conditions which correspond to the adiabatic vacuum,
\begin{eqnarray}
{\Omega}_{{\bf k}s} (\tau_0) &=& \omega_{\bf k} (\tau_0) ,\nonumber \\
\dot{\Omega}_{{\bf k}s} (\tau_0) &=& \dot{\omega}_{\bf k} (\tau_0)
\end{eqnarray}
one can choose $N_{\pm}({\bf k}s) = F({\bf k}s) = 0$, thus
$N_{\pm}({\bf k}s,\tau) = \vert\beta ({\bf k}s ; \tau)\vert^2$.  Hence, the
time-dependent particle number defined by (\ref{s_A12}) with (\ref{Bogcoefb})
has the property of starting at zero at $\tau=0$ if the initial state is the
adiabatic vacuum state.  At late times, when electric fields go to zero it
approaches the usual out-state number operator.
Thus we can identify the phase-space density as
\begin{equation}
{\tilde f}( k_{\eta}, {\bf k}_{\perp},\tau) \equiv {\frac {dN}   {d\eta
\,dk_{\eta} \,d{\bf k}_{\perp} d{\bf x}_{\perp} }}   =
|\beta( k_{\eta}, {\bf k}_{\perp},\tau)|^2.
\label{boost_betas}
\end{equation}

The quantity defined in this way from first principles of the microscopic
quantum theory agrees quite well (after coarse graining) with the single
particle distribution function $f( {\bf p}, \tau)$, obtained by solving the
Boltzmann-Vlasov eq. (\ref{boostBV}). From the single particle distribution one
can construct the Boltzmann entropy density in the comoving frame,
\begin{equation}
s(\tau) = -{1 \over \tau} \int {dk_{\eta} \,d{\bf k}_{\perp} \over (2\pi)^3}
\bigl\{ {\tilde f}\ln {\tilde f} + (1 - {\tilde f})\ln (1 - {\tilde f})
\bigr\}\ .
\label{ent2}
\end{equation}
If ${\tilde f}$ approaches a Fermi-Dirac equilibrium distribution at late
$\tau$, then this Boltzmann entropy will agree with the quantity (\ref{ent1})
defined by the energy-momentum tensor in Section 4.

\myappsection{Fluid rapidity distribution and particle rapidity
distribution}

In hydrodynamical models one has a purely phenomenological
description in terms of the collective variables---energy density, pressure
and hydrodynamic four-velocity.  Using a criterion for hadronization
such as those described by Landau \cite{Landau} and by Cooper, Frye,
and Schonberg \cite{CFS} one determines the particle spectra by making a
further assumption that at break-up the fluid velocity is equal to the particle
velocity. In our field theory treatment no such further assumption is needed as
long as boost invariance holds and we determine the particle spectrum
along a surface of constant $\tau$\@. In this appendix we will show the
equivalence  $dN/d\eta = dN/dy$.

One ingredient in the proof is the fact that the transformation of coordinates
to the $(\eta, \tau)$ system is a transformation to a local frame that moves
with constant velocity $\tanh\eta$ with respect to the Minkowski
center-of-mass frame, i.e., the comoving frame is not accelerated with respect
to the Minkowski frame. Because of this, the total number of particles counted
in that frame is the same as the number of particles in the Minkowski frame of
reference. The second ingredient is that in order for the phase-space
volume to be preserved under our coordinate transformation, we
need to ensure that the transformation in phase space is
canonical in the classical sense of preserving Poisson brackets.

Consider then the coordinate transformation
\begin{eqnarray}
 \tau = (t^2-z^2)^{1/2}  \qquad \,
\eta = \frac{1}{2} \ln \left({{t+z} \over{t-z}} \right) \nonumber
\\
p_{\tau}= Et/ \tau -p z/ \tau \quad \, p_{\eta}  = -Ez + tp.
\label{boost_4trans}
\end{eqnarray}
The Poisson bracket is defined as
\begin{equation}
\{A,B \} =  {\partial A \over \partial p}{\partial B \over
\partial
x}  -
{\partial A \over \partial E}{\partial B \over \partial t}-
{\partial B \over \partial p}{\partial A \over \partial x}+
{\partial A \over \partial t}{\partial B \over \partial E} \,.
\label{boost_poisson}
\end{equation}
The Poisson brackets of these quantities are
\begin{eqnarray}
\{\tau,\eta \} = 0, \quad \{p_{\eta},p_{\tau} \} = 0,\nonumber \\
\{p_{\tau}, \tau \} = -1, \quad  \{p_{\eta},\eta \} = 1 .
\label{boost_4anticom}
\end{eqnarray}
We see that the above transformation is canonical.

The phase-space density of
particles can be derived as shown in Appendix A, and
it is  found to be $\eta$-independent.
In order to obtain the rapidity distribution,
we change variables from $(\eta, k_{\eta})$ to
$(z, y)$,
where $y$ is the particle rapidity,
\begin{equation}
y = \frac {1}{2}  \ln \left ({{E+ k_{z}} \over  {E- k_{z}}
}\right ) .
\label{boost_y}
\end{equation}
Thus we have
\begin{eqnarray}
{\frac {dN}   {d\eta\,dk_{\eta}\ d{\bf k}_{\perp}d{\bf
x}_{\perp}}}&=&
J\,\frac {dN}{dy\,dz\,d{\bf k}_{\perp} d{\bf x}_{\perp}}\,,
\label{boost_J}
\end{eqnarray}
where the Jacobian is
\begin{equation}
 J ^{-1}  = \left| \matrix{
 {\partial k_{\eta}}/{\partial y}&{\partial
k_{\eta}}/{\partial z} \cr
{\partial \eta}/{\partial y}&
{\partial \eta}/{\partial z} \cr}\right|
= \frac {\partial k_{\eta}}{\partial y}
\frac{\partial\eta}{\partial z} .
\label{boost_Jinv}
\end{equation}
$p_{\tau}$ and $p_{\eta} $ can be rewritten as
\begin{eqnarray}
p_{\tau} &=& m_{\perp} \cosh(\eta-y) \nonumber \\
p_{\eta} &=& -\tau m_{\perp} \sinh(\eta-y)\,.
\label{boost_ptaueta}
\end{eqnarray}
The particle spectrum is calculated at a fixed value of $\tau$,
so
$\eta = \sinh^{-1}({z}/{\tau}) = \eta(z)$\@.
Thus functionally at fixed $\tau$ we have
\begin{equation}
p_{\eta}+eA_{\eta}\equiv k_{\eta} = k_{\eta}(\eta(z) -y)\,.
\label{boost_keta}
\end{equation}
The chain rule then gives
\begin{eqnarray}
\left.\frac {\partial k_{\eta}}{\partial z}\right|_{\tau} =
  \frac {\partial
k_{\eta}}{\partial\eta}\frac{\partial\eta}{\partial z}
 = - \frac {\partial k_{\eta}}{\partial y}
\frac{\partial\eta}{\partial z}\,.
\label{boost_chain}
\end{eqnarray}
At constant $\tau$, then, $|J| =dz/dk_\eta$,
which leads to the desired result
\begin{equation}
\frac{dN}{dy} = \frac{dN} {d\eta}\,.
\label{boost_proof}
\end{equation}
Since  the right-hand side of (\ref{boost_betas}) is $\eta$
independent,
$dN/d\eta$ is flat in $\eta$\@. From (\ref{boost_proof}) we
conclude that
the distribution $dN/dy$ is flat, as expected.

\myappsection{Scalar electrodynamics in conformal coordinates}
Since $\tau=0$ is a singular point of our equations, we find it
convenient to introduce the conformal time coordinate $u$ via\footnote{In
the radial Schrodinger equation a singularity at $r=0$ prevents
straightforward application of the WKB method.
The transformation $r=r_0\exp u$
maps the singularity from the origin to $-\infty$ and
enables one to use the WKB approximation in terms of the new
variable $u$ \cite{Morse}.
Our situation is different, because
in the vicinity of the singular point $\tau=0$ we
can still use the adiabatic expansion in terms of $\tau$; the
variable
$u$ is still helpful in avoiding numerical difficulties.}
\begin{equation}
m\tau=e^u\ .
\end{equation}
In (1+1) dimensions the line element reads
\begin{eqnarray}
ds^2=-dt^2+dz^2=\frac{e^{2u}}{m^2}(-du^2+d\eta^2)\,.
\label{boost_u_line_element}
\end{eqnarray}
The transformation from the Minkowski $t,z$ coordinates
to the  Kasner $u,\eta$ coordinates is conformal.
We shall refer to $u$ as the conformal proper time.\footnote{
For a general Kasner metric the conformal time is defined as
$\eta_{conf}\equiv\int^{\tau}[(-g)^{1/2}]^{-1/3}$, where $g$ is
the determinant of (\ref{boost_metric}).}

Instead of expanding the field $\chi$ as in (\ref{boost_field})
we expand the field $\phi$ with the mode functions $g_k=f_k/\sqrt{\tau}$,
which satisfy
\begin{equation}
\frac{d^2 g_k}{d\tau^2}+\frac{1}{\tau}\frac{dg_k}{d\tau}
+\left[m_{\perp}^2
+\frac{(k_{\eta}-eA(\tau))^2}{\tau^2}\right]g_k(\tau)=0\,,
\end{equation}
where $m_{\perp}^2\equiv k_{\perp}^2+m^2$\@.
In terms of $u$ the mode equation is
\begin{equation}
\frac{d^2g_k}{du^2}+w_k^2(u) g_k(u)=0\,,
\end{equation}
where
\begin{equation}
w_{k}^2(u)\equiv
\frac{m_{\perp}^2}{m^2}e^{2u}+(k_{\eta}-eA(u))^2\,.
\label{boost_gu2}
\end{equation}
We parametrize $g_k$ in a WKB-like ansatz,
\begin{eqnarray}
g_k(u)=\frac{1}{\sqrt{2W_k(u)}}\exp\left (-i\int^u W_k(u')du'
\right )\,,
\label{boost_uansatz}
\end{eqnarray}
and again the real mode functions $W_k$ satisfy the differential
equation
\begin{eqnarray}
\frac{1}{2}\frac{\ddot{W}_k}{W_k}-\frac{3}{4}\frac{\dot{W}_k^2}{W_k^2}
+W_k^2=w_k^2(u)\,,
\label{boost_uWKB}
\end{eqnarray}
where the dot now denotes differentiation with respect to $u$\@.
The Maxwell equation for an initial vacuum state is now
\begin{equation}
-\frac{dE}{du}=
j_{\eta}
\end{equation}
or
\begin{equation}
e^{-2u}\left(\frac{d^2 A_{\eta}}{du^2}
 -2\frac{dA_{\eta}}{du}\right)=
e\int[d{\bf k}]\,\frac{k_{\eta}-eA_{\eta}(u)}{W_k(u)}\,.
\label{boost_uMax}
\end{equation}

Performing the adiabatic expansion of $W_k(u)$ for large $k$
we find
\begin{equation}
e^{-2u}\left(\frac{d^2 A_{\eta}}{du^2}
-2\frac{dA}{du}\right)=
-e^{-2u}\left(\frac{d^2 A_{\eta}}{du^2}
-2\frac{dA_{\eta}}{du}\right)
e^2\delta e^2
+({\rm finite\  terms}) \,,
\end{equation}
where
\begin{equation}
\delta e^2 = \frac{1}{24\pi^2}\ln\Lambda/m
\end{equation}
in (3+1) dimensions, and
\begin{equation}
\delta e^2 =\frac{1}{12\pi m^2}
\end{equation}
in (1+1) dimensions, as expected.  Note that the renormalization
procedure introduces no additional interactions, in contrast to
(\ref{boost_1Ddeltae}) in $(\eta,\,\tau)$ coordinates.

\newpage

\centerline{\Large \bf Acknowledgments}
\vskip\baselineskip
This work was partially supported by the German-Israel
Foundation. Further support was provided by
the Ne'eman Chair in Theoretical Nuclear Physics at Tel Aviv
University. The work of B. S. was supported
by a Wolfson Research Award administered by the Israel Academy of
Sciences and Humanities. Y.K. thanks the Theoretical Division of Los Alamos
National Laboratory for their hospitality.
J.M.E. thanks the Institute for Theoretical Physics
of the University of Frankfurt, and its director, Professor
Walter Greiner, for their hospitality, and
the Alexander von Humboldt-Stiftung for partial support of this
work. Finally, the authors gratefully acknowledge the Advanced Computing
Laboratory at Los Alamos for the use of their computers and facilities, and
Pablo Tamayo for translating the code onto the Connection Machine.

\newpage

\centerline{\bf FIGURE CAPTIONS}
\vskip\baselineskip
\parindent0pt
\parskip14pt

FIGURE 1. Conformal proper time evolution of (a) the gauge
field $A(u)$, (b) electric field $ E(u)$, and (c) current
${\jmath}_{\eta}(u)$, for scalar particles in dimensionless units
(\ref{boost_dimensionless}). The initial conditions are that of adiabatic
vacuum with respect to conformal $u$ time at $u=-2$ with
initial electric field $ E (u=-2)=4.0$

FIGURE 2. Conformal proper time evolution of electric
field (a) $ E(u)$, and (b) scalar particle current $ {\jmath}_{\eta}(u)$ with
the same initial conditions as in Fig.~ 1 (solid lines) compared to solution
of the Boltzmann-Vlasov equation (dashed line). (c) and (d) are the same as (a)
and (b) but for initial adiabatic vacuum conditions at $u=0$.

FIGURE 3. Proper time evolution of the system of (a) electric field
$ E (\tau)$, and (b) fermion current $ {\jmath}_{\eta}(\tau)$, for initial
conditions at $\tau=1$ with initial electric field $E (\tau=1)=4.0$

FIGURE 4. Proper time evolution of  $\tau\epsilon(\tau)$ for fermions.

FIGURE 5. Proper time evolution of  $p/\epsilon$ for fermions.

FIGURE 6. Proper time evolution of $dN/ d\eta $ for fermions.

FIGURE 7. Proper time evolution of $\tau\epsilon/(dN/ d\eta)$ for fermions.

FIGURE 8. Proper time evolution of Boltzmann entropy density $s$, multiplied by
$\tau$ for fermions.

FIGURE 9. Proper time evolution of the effective hydrodynamical ``temperature"
for fermions.

\end{document}